\documentclass[manuscript,nonacm]{acmart}
\usepackage{xspace}
\newcommand{\system}{\textsc{PersonaCite}\xspace} 

\AtBeginDocument{%
  }

\acmConference[CHI EA '26]{CHI Conference on Human Factors in Computing Systems Extended Abstracts}{April 13--17, 2026}{Barcelona, Spain}

\begin{document}
\title[\system: VoC-Grounded Interviewable Verifiable Agentic Synthetic AI Personas for User Research]{\system: VoC-Grounded Interviewable Agentic Synthetic AI Personas for Verifiable User and Design Research}
\subtitle{Preprint — Under review.}

\author{Mario Truss}
\email{mtruss@adobe.com}
\affiliation{%
  \institution{Adobe}
  \country{Germany}
}
\renewcommand{\shortauthors}{Truss}

\begin{abstract}
LLM-based and agent-based synthetic personas are increasingly used in design and product decision-making, yet prior work shows that prompt-based personas often produce persuasive but unverifiable responses that obscure their evidentiary basis. We present \system, an agentic system that reframes AI personas as evidence-bounded research instruments through retrieval-augmented interaction. Unlike prior approaches that rely on prompt-based roleplaying, \system{} retrieves actual voice-of-customer artifacts during each conversation turn, constrains responses to retrieved evidence, explicitly abstains when evidence is missing, and provides response-level source attribution. Through semi-structured interviews and deployment study with 14 industry experts, we identify preliminary findings on perceived benefits, validity concerns, and design tensions, and propose Persona Provenance Cards as a documentation pattern for responsible AI persona use in human-centered design workflows.
\end{abstract}

\ccsdesc[500]{Human-centered computing~HCI design and evaluation methods}
\ccsdesc[500]{Human-centered computing~User centered design}
\ccsdesc[300]{Computing methodologies~Natural language generation}
\ccsdesc[300]{Computing methodologies~Neural networks}

\keywords{AI personas, synthetic users, agentic systems, data grounding, reaction simulation, VoC data, hallucination mitigation}

\begin{teaserfigure}
  \includegraphics[width=\textwidth]{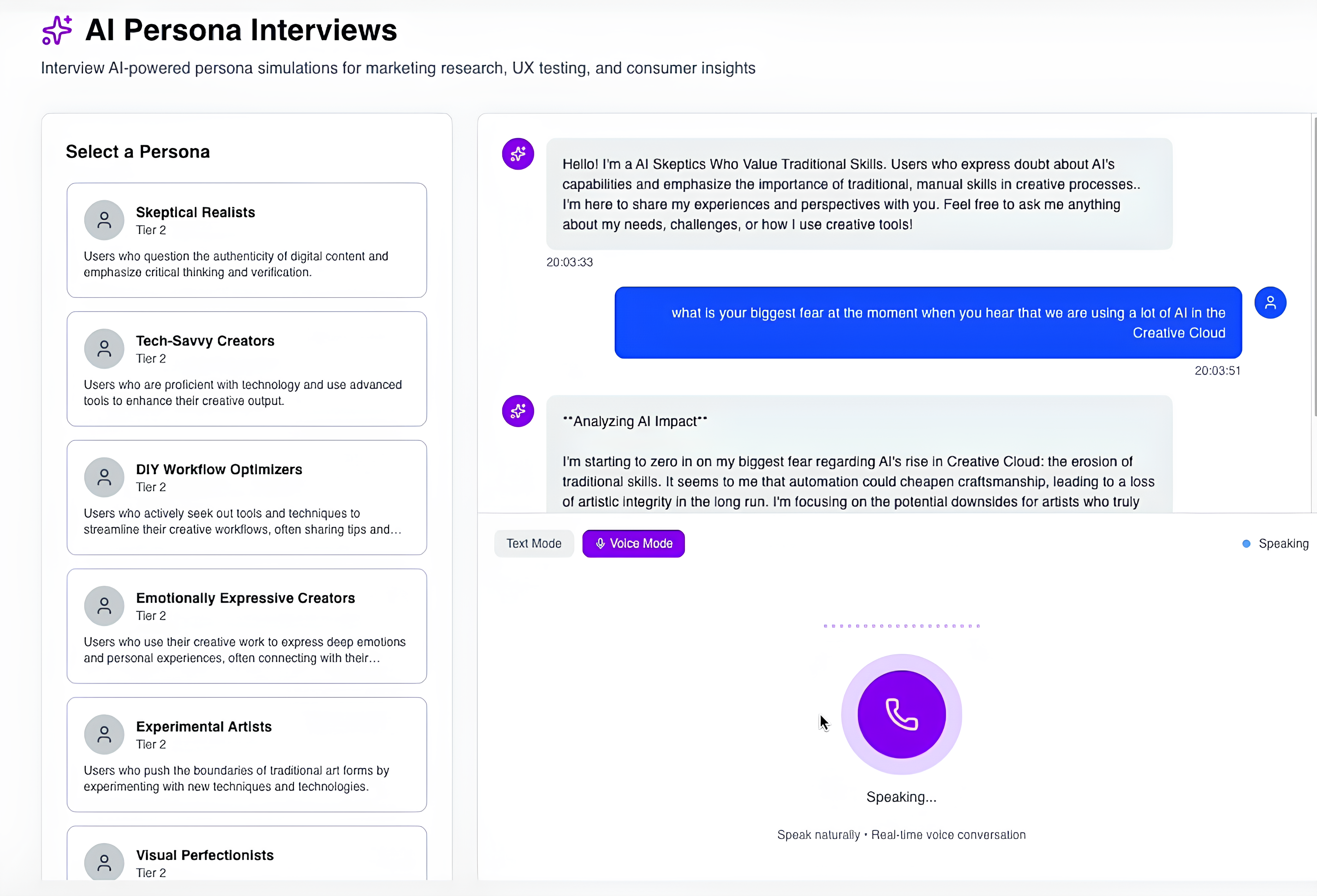}
  \caption{Interviewing a data-grounded synthetic persona. Personas respond only when supported by evidence and explicitly acknowledge gaps in available data. The examples shows a discussion between a designer and the synthetic person. It's an illustrative brand and was chosen, because it's widely known in the HCI and design domain, plus there is a lot of public data available. \system{} is a research prototype and is not a commercial product or service and results are illustrative.}
  \Description{Screenshot of a conversational interface showing a grounded synthetic persona interview.}
  \label{fig:teaser}
\end{teaserfigure}

\maketitle

\section{Introduction}

Interactive synthetic or simulated AI personas promise rapid access to user perspectives compared to static persona artifacts, that might be rooted purely in ideation by the researcher or based on limited and potentially biased manual research results. As natural language processing innovations enable systematic analysis of consumer behavior and user preferences \cite{Feuerriegel2025NLP}, LLM-based personas are increasingly used in design and product decision-making. However, recent work demonstrates that LLM-based personas are often weakly grounded, inconsistent and prone to hallucinating plausible yet unverifiable user opinions \cite{Salminen2025PersonaPrompts,Amin2025IJHCS_AIPersonas,hamalainenEvaluatingLargeLanguage2023}. These limitations are risks for the quality of decision making through over reliance on predictions, which is particularly problematic when personas are used in early design or strategic contexts.

Recent research underlines the importance of grounding such personas in empirical user data and positioning AI personas as complementary research instruments rather than substitutes for direct user engagement \cite{Jung2025PersonaCraft,Salminen2025IntelligentPersonas}. However, existing interactive personas rely on prompt-based roleplaying or pre-computed statistical summaries \cite{shinUnderstandingHumanAIWorkflows2024}, lacking systematic evidence retrieval and verification during conversation. 

To address these shortcomings, we introduce \system, a system that operationalizes grounding through \textit{real-time evidence retrieval}: during each interaction, the system retrieves actual VoC artifacts, constrains responses to retrieved evidence, explicitly abstains when data is insufficient, and provides response-level source attribution, which addresses key validity concerns identified in recent CHI research on LLM-based simulation \cite{wangLargeLanguageModels2025,pangUnderstandingLLMificationCHI2025}.

\section{Related Work}

\noindent\textbf{Data-Grounded Persona Generation.} Synthetic persona methods grounded in social science data, such as surveys, census data, and large-scale behavioral datasets, improve representativeness compared to manually authored personas \cite{Jung2025PersonaCraft,Amin2025PersonaGeneration}, but are typically static artifacts that do not support interactive interrogation or real-time reaction testing.

\vspace{0.5em}\noindent\textbf{Interactive LLM-Based Synthetic and Simulated Personas.} The evolution from static to conversational personas \cite{Kaate2025DIS} demonstrates creative and exploratory value \cite{Park2023,Sun2025PersonaL}, but such systems largely rely on prompt engineering and lack systematic verification mechanisms \cite{Salminen2025PersonaPrompts}. Prior evaluations underline inconsistency, shallow role adherence, and limited transparency regarding data provenance \cite{Amin2025IJHCS_AIPersonas}. Related work on ``synthetic users'' \cite{guSyntheticUsersInsights2025} and writer-defined AI personas \cite{benharrakWriterDefinedAIPersonas2024} demonstrates value while highlighting consistency challenges. Work on grounding generative agents in real individual data \cite{Park2023} provides evidence for empirical grounding in persona simulation.

\vspace{0.5em}\noindent\textbf{Limitations of LLM-Based Simulation.} Recent CHI research reveals critical validity concerns with LLM-generated personas and synthetic data. Studies show that LLMs can misportray and flatten identity groups \cite{wangLargeLanguageModels2025}, produce believable but potentially unreliable synthetic research data \cite{hamalainenEvaluatingLargeLanguage2023}, and reveal reproducibility concerns when used as simulated users \cite{pangUnderstandingLLMificationCHI2025}. Work on human-AI workflows for persona generation demonstrates improved results through collaborative approaches that combine human expertise with LLM capabilities \cite{shinUnderstandingHumanAIWorkflows2024}. Further, human-centered AI research emphasizes transparency, accountability, and appropriate trust calibration thorugh explicit evidence constraints and provenance mechanisms \cite{Akhmetov2025CSCW}. Critical perspectives from UX research highlight that synthetic users struggle to capture unpredictable human behavior, may reflect biases in training data (including WEIRD cultural biases), and cannot fully replace the qualitative depth gained from observing real users \cite{russellChallengesSyntheticUsers2025}. These limitations motivate \system's focus on explicit evidence constraints, abstention behavior, and transparent communication of knowledge boundaries. 

\section{Methodology}

We conducted a formative evaluation study as part of an internal innovation project spanning 3 month, aiming to understand perceived value, limitations, and validity concerns of grounded, interviewable personas in real design workflows.

\vspace{0.5em}\noindent\textbf{Research Methods.} The study employed three complementary methods: (1) \textit{Formative evaluation of prototypes} through iterative testing and refinement, (2) \textit{semi-structured interviews} with domain experts to gather feedback on system components and interaction patterns, and (3) \textit{longitudinal collaboration} with regular cadences for gathering expert feedback.

\vspace{0.5em}\noindent\textbf{Participants and Study Structure.} Fourteen experts from UX research, product management, design, and AI strategy participated via Teams calls (recorded for analysis, see Appendix for participant info), ranging from 30 minutes to one hour. The iterative design process allowed us to refine system features based on ongoing feedback.

\vspace{0.5em}\noindent\textbf{Evaluation Context.} We employed a user-centered design approach, iteratively co-designing the system with expert stakeholders to ensure alignment with real-world design workflows and validity requirements. Participants evaluated the system through exploratory interaction scenarios and by testing design stimuli including feature ideas, mockups, problem statements, social media posts, and landing pages.

\vspace{0.5em}\noindent\textbf{Data Collection.} Feedback was collected through recorded discussions, guided conversations about interaction scenarios, and reflective synthesis sessions. Participants evaluated the system's utility, trustworthiness, and validity concerns. This study provides formative insights; a larger-scale validation study remains planned.

\section{System Design}

\begin{figure}[H]
  \centering
  \includegraphics[width=\linewidth]{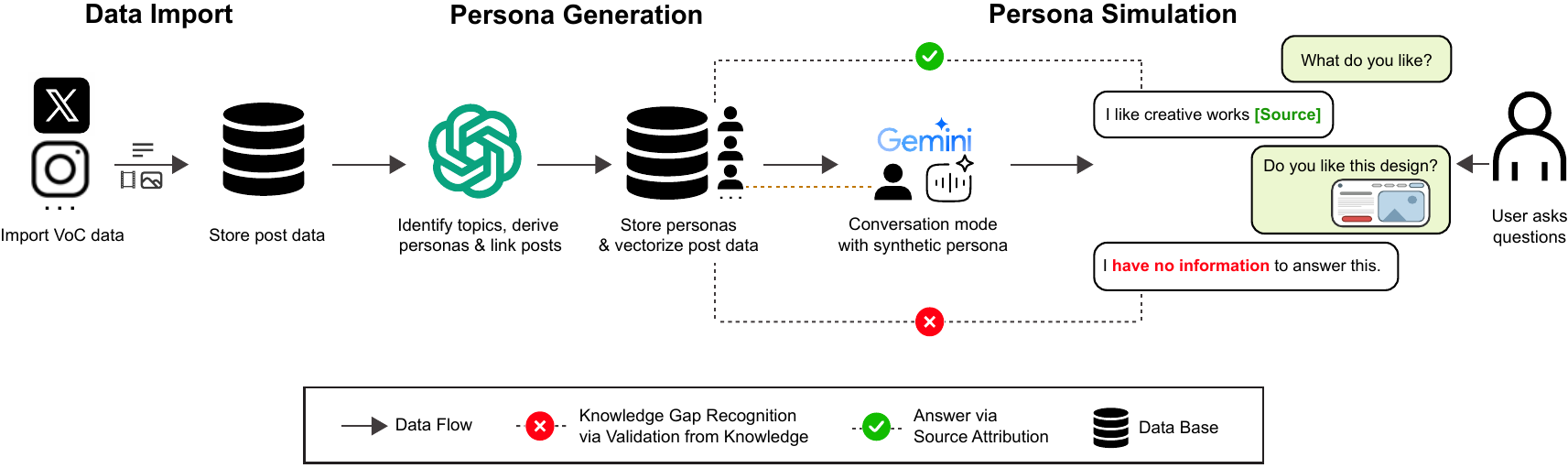}
  \caption{\system{} architecture. VoC data from social media and other channels is imported and stored. Multimodal AI identifies topics and derives personas, storing personas alongside vectorized post data. Persona interaction and reaction simulation with response-level source attribution is enabled via data-grounded conversational AI, with explicit knowledge gap recognition when evidence is insufficient.}
  \Description{Architecture diagram showing data ingestion, persona discovery, aggregation, and conversational interaction.}
  \label{fig:architecture}
\end{figure}

\system{} follows an agentic architecture implemented using Python (Pydantic for validation), AI.SDK and Next.js. The system leverages Gemini for conversational dialogue and GPT-4O as the core processing engine. Multimodal VoC data (text, images, video transcripts) from social media and other channels are imported and stored, then processed to identify topics and derive personas. The derived personas and vectorized post data are stored together, enabling persona simulation through a conversational LLM constrained to respond exclusively based on stored evidence with response-level source attribution and explicit knowledge gap acknowledgment. 

\vspace{0.5em}\noindent\textbf{Content Engineering.} Our approach builds on \textit{agentic context engineering (ACE)} \cite{zhangAgenticContextEngineering2025}, which has demonstrated that retrieving and providing the right evidence as context improves both response quality and grounding while avoiding brevity bias and context collapse. From what we were able to find, this is novel approach and was not used before.

\vspace{0.5em}\noindent\textbf{Interaction Modes.} The system supports two interaction modes: (1) \textit{persona interviews} for exploratory inquiry, and (2) \textit{reaction simulation}, where personas are presented with concrete design stimuli such as feature descriptions, interface mockups, or messaging concepts.

\vspace{0.5em}\noindent\textbf{Explicit Gap Acknowledgment.} When insufficient evidence exists, personas explicitly abstain and communicate topic coverage limits rather than generating speculative responses, directly addressing known validity risks in AI persona research \cite{Amin2025IJHCS_AIPersonas,Sethi2025WhenAIWrites}.

\vspace{0.5em}\noindent\textbf{Source Attribution.} Persona responses are accompanied by post-hoc conversation summaries linking each claim to underlying VoC artifacts, enabling verification, traceability, and reuse of verbatim user language, which stakeholders appreciated.

\section{Findings \& Discussion}

\system{} advances prior persona systems through three key mechanisms that shift grounding from creation-time to interaction-time. While existing approaches use data to generate persona descriptions or pre-computed statistics \cite{Jung2025PersonaCraft,shinUnderstandingHumanAIWorkflows2024}, conversational responses rely on prompt-based roleplaying that can hallucinate beyond available evidence \cite{Salminen2025PersonaPrompts,hamalainenEvaluatingLargeLanguage2023}. \system's innovation is \textit{retrieval-augmented persona simulation}: (1) during each conversation turn, the system retrieves actual VoC artifacts from the evidence base; (2) LLM responses are constrained to only claim what retrieved evidence supports; (3) when insufficient evidence exists, the system explicitly abstains rather than generating plausible speculation. This moves beyond persuasive simulation toward verifiable research instruments \cite{Amin2025PersonaGeneration,Batzneretal2025}, addressing critical validity concerns about identity misrepresentation \cite{wangLargeLanguageModels2025} and synthetic data reliability \cite{pangUnderstandingLLMificationCHI2025}. We reframe AI personas as interactive archives of empirical evidence, where personas simulation serves as exploratory sensemaking rather than high-fidelity prediction, similarly described before  \cite{Schmidt2024Interactions}.

\vspace{0.5em}\noindent\textbf{Reaction Simulation as a Design Accelerator.} Participants consistently highlighted reaction simulation as valuable for early-stage design exploration, rapid iteration, and stakeholder alignment. Testing multiple design concepts without waiting for user recruitment was seen as a significant workflow advantage.

\begin{quote}
\emph{``Being able to ask a persona how users would react before anything is built fundamentally changes how fast we can iterate. Some things you simply can't test before and would prevent potential negative backlash''} (P6)
\end{quote}

Participants mentioned that reaction simulation enabled rapid hypothesis testing and early identification of user concerns, although they emphasized that this should complementary rather than replacement for direct user engagement. \textit{Will AI replace user research?} Our findings suggest the answer is no. Grounded personas are a tool that complements, rather than substitutes for, direct user engagement. They excel at rapid exploration and hypothesis testing when user access is limited, but cannot capture the nuanced, contextual insights from observing real users. Grounded personas address persistent challenges in user research workflows: Traditional research is time-consuming and costly, and does not always yield actionable insights. \system{} analysis enables personas to be grounded in diverse VoC channels (social media, support tickets, user-generated content), as it's readily available and allows maintaining traceability to authentic user perspectives.

\vspace{0.5em}\noindent\textbf{Transparency, Traceability, and Trust.} A central finding was the critical importance of understanding the evidentiary basis for persona responses. 

\begin{quote}
\emph{``This type of simulation is almost like a form reputation management and business intelligence, not just a design tool but I need to know that it's true. When we say something is proven, we need to be extra cautious to not lose trust in case of failure''} (P4)
\end{quote}

Participants expressed strong concerns about LLM hallucinations and emphasized the need for confidence scores, source traces, and citations linking claims back to original VoC artifacts. This feedback directly influenced our final design, which includes source attribution at the response-level and explicit acknowledgment of knowledge gaps. Our preliminary findings reveal that trust in AI personas fundamentally depends on transparency about data provenance and response generation. Participants' requests for confidence scores, source traces, and citations reflect broader concerns about LLM reliability. The challenge of distinguishing between individual opinions and generalizable patterns in noisy internet data further underscores the need for clear documentation of data quality and segment representativeness. Participants also highlighted the challenge of differentiating between individual opinions and generalizable patterns in noisy internet data, with several expressing interest in causal inference capabilities (future work). Recent large-scale studies of digital twins reveal that while such systems can capture relative differences in individuals, they struggle to adequately represent unique individual judgments \cite{pengMegaStudyDigitalTwins2025}. These findings showcase the need for grounded personas. Notably, participants particularly appreciated the source attribution and explicit knowledge gap acknowledgment – features directly enabled by our agentic context engineering approach \cite{zhangAgenticContextEngineering2025}. This suggests that retrieval-augmented architectures align well with practitioner needs for verifiable, transparent AI research tools.

\vspace{0.5em}\noindent\textbf{Grounding Improves Trust but Not Certainty.} Explicit grounding, abstention behavior, and source attribution increased perceived responsibility and appropriate trust calibration. However, participants remained cautious about subtle extrapolation beyond available evidence and wanted more granular transparency about data quality, segment representativeness, and potential biases.

\vspace{0.5em}\noindent\textbf{Validity as a Design Variable, Not a Binary Criterion.} Rather than rejecting the system, experts framed validity concerns as a design requirement: The need for transparent positioning, documentation of limitations, and complementary use alongside traditional user research. Participants acknowledged that while grounded personas reduce risks compared to ungrounded LLM personas, they still require careful interpretation and should be positioned as exploratory tools rather than definitive user research. Our findings suggest a broader provocation for HCI: Rather than asking whether AI personas are ``valid,'' we should ask how validity is designed, communicated, and constrained in interactive systems. Participants did not reject \system{} due to validity concerns; instead, they treated validity as negotiable through transparency, abstention, and documentation. This reframes validity from a binary evaluation criterion into a design variable shaped through interface mechanisms, provenance disclosures, and explicit scoping. The central risk of AI personas is not inaccuracy, but implicit limitations. Making these limits visible may be more consequential for responsible design practice than improving predictive fidelity alone.

\vspace{0.5em}\noindent\textbf{Persona Provenance Cards.} To support responsible deployment and mitigate over-trust, we propose \textit{Persona Provenance Cards} as a standard documentation pattern, extending model cards \cite{Mitchelletal2018} and datasheets \cite{Gebruetal2018} to interactive persona systems. These cards should document: \textbf{Data Provenance} (VoC channels, social platforms, collection methods, temporal range), \textbf{Model Specifications} (underlying LLM/MMM usage and potential risks), \textbf{Segment Metrics} (number of users and messages grounding each persona), and \textbf{Topic Coverage} (documented areas of data availability and evidentiary gaps).

\vspace{0.5em}\noindent\textbf{Limitations and Future Work.} This study provides formative insights from expert stakeholders in a design sprint context. The findings represent anecdotal evidence from a limited participant group and may not generalize to broader populations. Critically, while participants appreciated the transparency and source attribution features, we did not validate the factual correctness or accuracy of persona responses against ground truth in a larger study. Future work should include systematic benchmark studies comparing \system{} against other persona approaches and human-created personas to empirically demonstrate improvements in accuracy, reliability, and validity beyond user perception. We plan to conduct a larger-scale validation study deploying \system{} with design teams in realistic project settings over multiple months, comparing persona-supported design decisions with traditional user research outcomes. We will investigate failure modes such as extrapolation errors and data sparsity impacts through controlled experiments with ground-truth comparisons. Additionally, we plan to extend the system with causal inference capabilities to solidify claim via showcasing underlying drivers, their effect size and causal relationships in VoC data. This would allow moving beyond correlation to support more actionable design insights. Future work should further explore adding additional user context might help, such as screen sharing capabilities for live design environment interaction.

\section{Conclusion}

\system{} demonstrates how VoC-data grounding, combined with evidence retrieval, explicit abstention, and response-level provenance can operationalize responsible AI personas for human-centered design. Unlike prior approaches that use data to create persona descriptions or statistical summaries but rely on prompt-based roleplaying during interaction \cite{Jung2025PersonaCraft,shinUnderstandingHumanAIWorkflows2024}, \system{} retrieves actual VoC artifacts during each conversation turn and constrains responses to retrieved evidence. This addresses validity concerns about hallucination \cite{hamalainenEvaluatingLargeLanguage2023} and identity misrepresentation \cite{wangLargeLanguageModels2025} through systematic verification rather than prompt engineering. By reframing validity as a design variable shaped through transparency mechanisms rather than a binary evaluation criterion, we contribute an approach to building trustworthy AI research instruments. Our proposed Persona Provenance Cards provide a concrete documentation pattern for responsible deployment. Grounded personas support early-stage exploration when transparently documented and appropriately scoped, complementing rather than replacing user research. As more GenAI and agentic AI systems become part of the designers' daily lives and workflows, explicitly designing for verifiability user research insights becomes essential for responsible innovation.


\bibliographystyle{ACM-Reference-Format}
\bibliography{bib}

\section{Appendix}
\appendix
\begin{table}[h]
  \centering
  \caption{Expert consultation participants who provided feedback on \system}
  \label{tab:expert-participants}
  \begin{tabular}{llll}
  \hline
  \textbf{Participant} & \textbf{Role} & \textbf{Level} & \textbf{Gender} \\
  \hline
  P1 & Social Media & Head & Male \\
  P2 & UX Research & Senior & Male \\
  P3 & Community Management & Director & Female \\
  P4 & Social Intelligence & Senior & Female \\
  P5 & Digital Experience & Director & Male \\
  P6 & Product Management & Director & Male \\
  P7 & Strategy & Senior & Male \\
  P8 & Experience Design & Director & Female \\
  P9 & Design & Director & Female \\
  P10 & Technical Consulting & Senior & Male \\
  P11 & Technical Consulting & Consultant & Male \\
  P12 & Product Management (AI/ML) & Senior & Male \\
  P13 & Customer Success & Manager & Female \\
  P14 & AI Product & Head & Female \\
  \hline
  \end{tabular}
\end{table}  

\end{document}